\documentclass[aps,prb,twocolumn,superscriptaddress,showpacs,floatfix]{revtex4}

\usepackage{graphicx}
\usepackage{latexsym}
\usepackage{stmaryrd}
\usepackage{amsmath}
\usepackage{amssymb}
\usepackage{epstopdf}

\usepackage[colorlinks,linkcolor=red,anchorcolor=blue,citecolor=green]{hyperref}

\begin{document}

\title{Topological Anderson Insulator in electric circuits}

\author{Zhi-Qiang Zhang }
\affiliation{School of Physical Science and Technology, Soochow University, Suzhou, 215006, China}
\author{Bing-Lan Wu }
\affiliation{School of Physical Science and Technology, Soochow University, Suzhou, 215006, China}
\author{Juntao Song } \email{jtsong@hebtu.edu.cn}
\affiliation{Department of Physics and Hebei Advanced Thin Film Laboratory, Hebei Normal University,
Shijiazhuang, 050024, China}
\author{Hua Jiang}\email{jianghuaphy@suda.edu.cn}
\affiliation{School of Physical Science and Technology, Soochow University, Suzhou, 215006, China}
\affiliation{Institute for Advanced Study, Soochow University, Suzhou 215006, China}
\date{\today}

\begin{abstract}
Although topological Anderson insulator has been predicted in 2009, the lasting investigations of this disorder established nontrivial state results in only two experimental observations in cold atoms [Science, {\bf 362 },929 (2018)] and in photonic crystals [Nature, {\bf 560}, 461 (2018)] recently. In this paper, we study the topological Anderson transition in electric circuits. By arranging capacitor and inductor network, we construct a disordered Haldane model. Specially, the disorder is introduced by the grounding inductors with random inductance. Based on non-commutative geometry method and transport calculation, we confirm that the disorder in circuits can drive a transition from normal insulator to topological Anderson insulator. We also find the random inductance induced disorder possessing unique characters rather than Anderson disorder, therefore it leads to distinguishable features of topological Anderson transition in circuits. Different from other systems, the topological Anderson insulator in circuits can be detected by measuring the corresponding quantized transmission coefficient and edge state wavefunction due to mature microelectronic technology.
\end{abstract}

\maketitle

\section{introduction}
Topological state$\cite{TIkane,zhang,xue,weyl1,weyl2,HOTI}$ is one of the most fascinating research areas in the past decades for its exotic properties. Due to the urgent need of applications in low-dissipative  devices, the disorder effect in topological states has also been widely investigated$\cite{Li,Hua1,Chen1,Chen2,Chen3,song1,song2,WYJ,Flo1,Flo2,hug1,hug2,Nagaosa1,Nagaosa2,HOdis}$ with the emergence of a variety of materials. Most studies indicate topological  states are robust against weak disorder $\cite{Chen2,hug1,HOdis}$ and their nontrivial features disappear under strong disorder.  Surprisingly, disorder does not always destroy topological properties$\cite{Li,Hua1,Chen1,WYJ,Flo2,hug1}$. In $2009$, by studying the effect of Anderson disorder in HgTe/CdTe Quantum wells $\cite{BHZ1,BHZ2,Hua1,Li}$, an interesting transition where a normal insulator becomes a topological insulator were predicted. This means disorder can also establish a topological state, which was named as topological Anderson insulator (TAI) $\cite{Li}$.
Later, TAI was explained by self-consistent Born approximation method, where disorder renormalizes the Hamiltonian with changing band structure from normal to inverted $\cite{lilun}$. The TAI and related topological transition have generated intensive studies in the last ten years$\cite{Chen1,Chen3,song1,Flo1,Flo2,HMGUO,reveiwDisorder,2011,2013,2014}$. However, until more recently, TAI phase was  experimentally confirmed by two independent groups. They find the existence of TAI transition in cold atom $\cite{shiyan1}$ and photonic crystal systems $\cite{shiyan2}$, respectively.

Recently, topological states simulated in the electric circuits also attract great attentions in condensed matter physics \cite{LCrev1,LCrev2,u1h1,u1h2,LCwey1,LCwey2,LCTB,LCHO1,LCHO2,LCorbit,LCgreen,LCchern1,LCchern2,LCedge,LCkane}. The realization of $U(1)$ hopping phase $\cite{u1h1,u1h2}$ suggests the possibility of simulating the effect of magnetic field and the appearance of Hofstadter Butterfly $\cite{Hofsta}$ in circuits. Various topological states have also been implemented, such as quantum spin Hall states$\cite{u1h1}$, Weyl semimetals$\cite{weyl1,weyl2}$, higher order topological insulator $\cite{LCHO1}$ and quantum anomalous Hall states$\cite{LCchern1,LCchern2}$ $etc.$. Compared with cold atoms and photonic crystals, parameters in circuits is easier to be controlled and physical quantities are more convenient to be measured$\cite{LCrev1,LCrev2,LCHO1,LCgreen}$. Since wave function in circuits is the voltage of each nodes$\cite{LCrev1,LCrev2}$ and Green's function corresponds to the impedance, both the wave function and the Green's function can be directly observed$\cite{LCHO1}$. For example, based on the impedance and voltage measurement, the detection of band spectrum$\cite{LCgreen}$, curvatures $\cite{LCwey2}$, zero energy state$\cite{LCHO1,LCHO2}$ and even orbital angular momentum$\cite{orbit,LCorbit}$ have been realized. These results manifest promising potential of circuits simulations in topological states.

In this paper, we propose a feasible method to realize TAI in circuits. By using cross connection method we rebuilt the Haldane model with disorder in electronic system. The topological nature of Haldane model can be controlled by selected inductors and the disorder is introduced by random inductance induced on-site potential $E_U$. Then, the possibility of TAI in our model is checked with the help of Chern number and transmission coefficient calculation. We find TAI exists in circuits. Furthermore, we also find topological Anderson transition in circuits is not the same to condensed matter conditions. The feature comes from the special distribution of $E_U$ rather than that of Anderson disorder. Compare with cold atoms and photonic crystals, the wave function and the Green's function in circuits correspond to the voltage$\cite{LCrev1}$ and the impedance$\cite{LCHO1}$, which are both easier to be measured.  Therefore, we propose the detection of the wave function (edge states) and Chern number$\cite{Thouless}$ (quantized transmission coefficient $T=1$) as the smoking gun evidences for TAI in circuits.  In addition, because of essential topological properties, one can determine TAI in circuits with need of only one dirty sample.

The rest of this paper is organized as follows: In Sec.$\uppercase\expandafter{\romannumeral2}$, we show the construction of disordered Haldane model in electric circuits.  In Sec.$\uppercase\expandafter{\romannumeral3}$, we obtain the general condition of realization of TAI in circuit. In Sec.$\uppercase\expandafter{\romannumeral4}$, we study the difference of TAI behavior between random inductance induced disorder and Anderson disorder.  In  Sec.$\uppercase\expandafter{\romannumeral5}$, we present experimental detection of TAI in circuits. Finally,  a brief discussion and summary are presented in Sec.$\uppercase\expandafter{\romannumeral6}$.

\section{Theoretic model}
\begin{figure}
    \centering
	\includegraphics[width=0.5\textwidth]{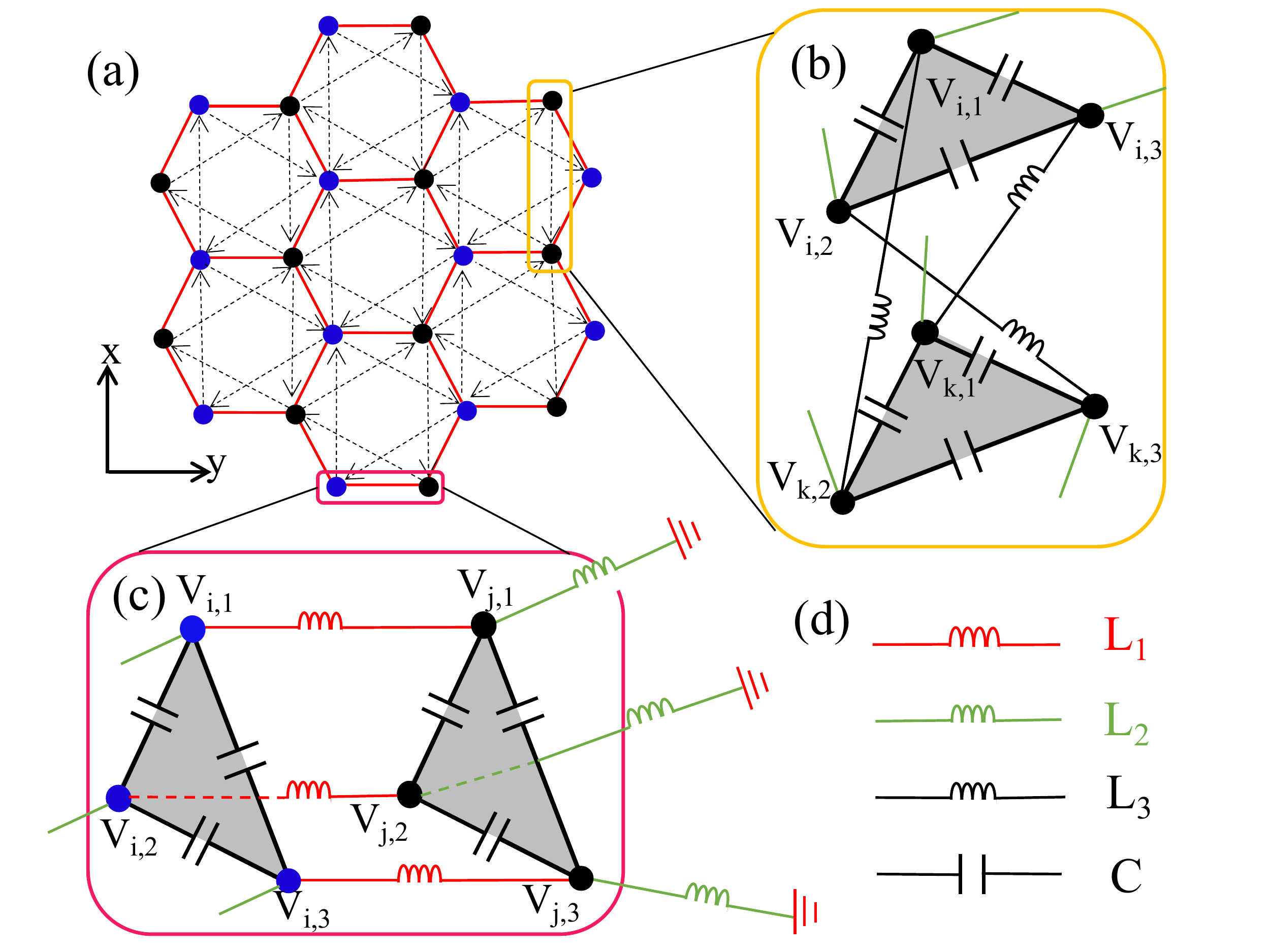}
	\caption{(Color online). Schematic plot of considered Haldane model with circuits. 	(a) illustrates honeycomb lattice structure of Haldane model,  where the red and black lines represent the nearest and next nearest hopping, respectively. The blue and black solid circles represent the A/B sublattice. Each site in Haldane model is constructed by the capacitor triangle in gray.
	(b) The two triangles correspond to the two sites as shown in (a). They are cross connected by black inductor with inductance $L_3$. This shows the realization of the next nearest hopping marked by orange square. (c)  is almost the same with (b) except the connection methods of red inductor with value $L_1$, which leads to the realization of nearest hopping in (a). The grounding inductor with value $L_2$ (in green color) is also demonstrated, which is important for realization of TAI. (d). All kinds of inductances and capacitances used in our system are marked. }\label{model}
\end{figure}
We begin with construct a Haldane model  with the help of circuits.
Haldane model is made of honeycomb lattice [see  Fig.~\ref{model}(a)], where nearest and next nearest hopping exist between A/B sublattice sites.
As illustrated in Fig.~\ref{model}(b) and (c), we firstly construct a triangle in gray color with three connected capacitors with the same capacitance $C$ and the voltages of all nodes are marked by $V_1,V_2,V_3$.  Every triangle could be considered as a site in honeycomb lattice and the color of triangle nodes represent the corresponding sublattice.
Next, three kind of inductances $L_1,L_3,L_2$ are used to simulate the nearest, next nearest hopping and on-site energy [see Fig.~\ref{model} (b)-(d)]. The red $L_1$ and black $L_3$ inductors stand for the nearest hopping and next-nearest hopping. Each node is also grounded by green inductor $L_2$. Fig.~\ref{model}(c) shows the realization of the nearest hopping in circuits where  $V_i$ of the left triangle is directly connected to  $V_j$ of the right triangle. The next nearest hopping shown in Fig.~\ref{model}(a) is realized in a different way.
We let  $V_i$ cross connected with  $V_{k}$, i.e. $\{V_{i,1}, V_{i,2},V_{i,3} \}$ is connected with $\{V_{k,2}, V_{k,3},V_{k,1} \}$, as shown in Fig.~\ref{model}(b). This cross connection method
can induce a $U(1)$ hopping with geometry phase, and has been used in previous studies to realize quantum spin Hall in circuits $\cite{u1h1,u1h2}$.

Following, we briefly introduce the relationship of Haldane model and the above circuit network. Based on Kirchhoff's  law\cite{LCHO1,LCrev1}, the AC current and the node voltage  in a circuit satisfy:
\begin{align}
\begin{split}
I_\alpha(\omega)&=i\omega^{-1}[-\sum_\beta \frac{1}{L_{\alpha\beta}}(V_\alpha-V_\beta)-\frac{1}{L_{\alpha}}(V_\alpha-0)\\
&+\sum_\gamma C_{\alpha\gamma }\omega^2(V_\alpha-V_\gamma)].
\end{split}
\end{align}
where $I_\alpha$ is the current of node $\alpha$, $V_\alpha$ is the voltage at node $\alpha$, $L_{\alpha\beta}$ is the inductance between nodes $\alpha$ and $\beta$, $L_\gamma$ is the inductance between nodes $\alpha$ and the ground, $C_{\alpha\gamma}$ is the capacitance between nodes $\alpha$ and $\gamma$. The summation are taken over all nodes $\beta$ and $\gamma$ which are connected with $\alpha$ by inductors and capacitors, respectively.

 In our model, the voltage and current for $i_{th}$ site can be written in vector form $\textbf{V}_{i}=[V_{i,1},V_{i,2},V_{i,3}]^T$ $\textbf{I}_{i}=[I_{i,1},I_{i,2},I_{i,3}]^T$. By applying Eq.(1) and $L_{\alpha\beta}=L_{1/3},L_\gamma=L_2,C_{\alpha\gamma}=C$, one obtain
\begin{align}
\begin{split}
\textbf{I}_{i}&=i\omega^{-1}[T_0\omega^2C-{\rm I_{3 \times 3}}/L_2]\textbf{V}_{i}\\
&-i\omega^{-1}[\sum_{\langle j\rangle}
\frac{\textbf{V}_{i}- \textbf{V}_{j}}{L_1}+\sum_{\langle\langle k\rangle\rangle}\frac{\delta \textbf{V}_{k}}{L_3}],
\end{split}
\end{align}
where
\begin{equation}
T_0=\left(
\begin{array}{ccc}
2 & -1 & -1\\
-1 & 2 & -1\\
-1 & -1 & 2\\
\end{array}
\right),
\end{equation}
and identity matrix $\rm I_{3 \times 3}$. The summation are taken over all the nearest $j$ and next nearest $k$ sites. Because of cross connection between site $i$ and $k$ [see Fig.~\ref{model}(b)], $\delta \textbf{V}_{k}= [V_{i,1}-V_{k,2},~V_{i,2}-V_{k,3},~V_{i,3}-V_{k,1} ]^{T}$. Then, Eq.(2) can be written as
\begin{align}
\begin{split}
\textbf{I}_{i}&=i\omega^{-1}[T_0\omega^2C-\rm I_{3 \times 3}(\frac{3}{L_1}+\frac{1}{L_2}+\frac{6}{L_3})]\textbf{V}_{i}\\
&+i\omega^{-1}[\sum_{\langle j\rangle}\frac{\rm I_{3 \times 3}\textbf{V}_{j}}{L_1}+\sum_{\langle\langle k\rangle\rangle}\frac{T_1\textbf{V}_{k}}{L_3}],
\end{split}
\end{align}
where $3/L_1$ and $6/L_3$ indicate three nearest and six next nearest neighbors $\cite{LCkane}$, and
\begin{equation}
T_1=\left(
\begin{array}{ccc}
0 & 1 & 0\\
0 & 0 & 1\\
1 & 0 & 0\\
\end{array}
\right).
\end{equation}
If there is no input current in circuit, $\textbf{I}_{i}=0$. By defining $E=T_0\omega^2C$, Eq.(4) has the following form:
\begin{align}
\begin{split}
E \textbf{V}_{i}&=(\frac{3}{L_1}+\frac{1}{L_2}+\frac{6}{L_3}) \textbf{V}_{i}- \\
&\sum_{\langle j \rangle} \frac{\rm I_{3\times 3}}{L_2}\textbf{V}_{j}-\sum_{\langle\langle k \rangle\rangle}\frac{T_1}{L_3} \textbf{V}_{k}.
\end{split}
\end{align}
If one analogizes $V$ with wavefunction, the above equation is close to the eigenequation for Haldane model with  $\frac{\rm I_{3\times 3}}{L_1}$ ($\frac{T_1}{L_3}$) the nearest (next nearest) hopping. The major difference comes from that $E$ is not diagonalized. In order to simulate Haldane model physics, we then perform unitary transformation with matrix$\cite{u1h1,u1h2}$:

\begin{figure}
    \centering
	\includegraphics[width=0.45\textwidth]{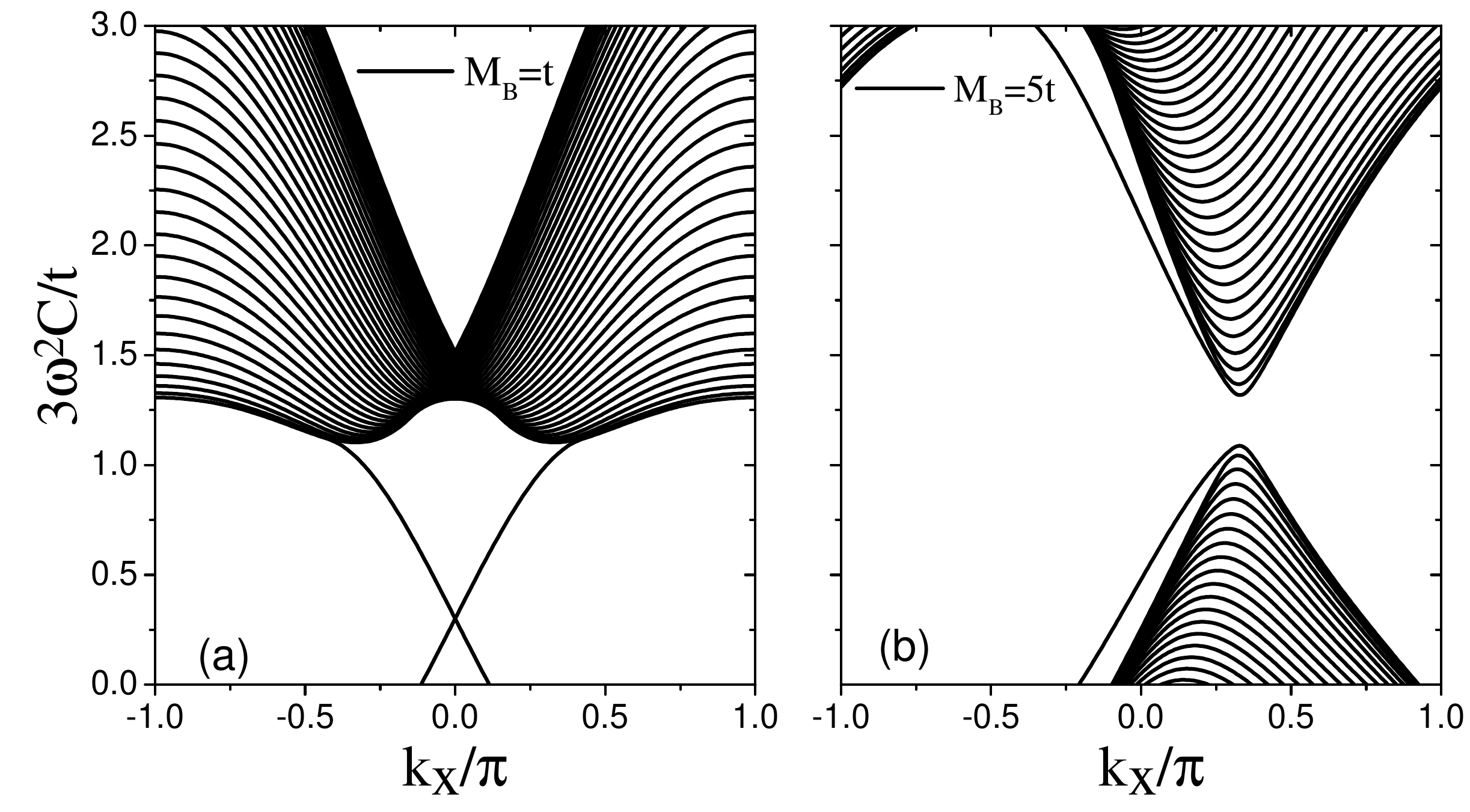}
	\caption{(Color online).  The band structures for electric circuits.(a) and (b) are plotted under different on-site potential for site B with $M_B=t$ and $5t$, respectively. The other parameters are set as $t=1$, $t_2=0.2t$, $\Delta=0.5$ and $M_A=t$. }\label{band}
\end{figure}

\begin{equation}
F=\frac{1}{\sqrt3}\left(
\begin{array}{ccc}
1 & 1 & 1\\
1 & e^{i2\pi/3} & e^{i4\pi/3}\\
1 & e^{i4\pi/3} & e^{i8\pi/3}\\
\end{array}
\right).
\end{equation}
Because $[T_0, T_1]=0$, $F$ can simultaneously diagonalize $T_0$ and $T_1$ into  $T_0^*=F^\dagger T_0F=diag[0,~3,~3]$ and $T_1^*=F^\dagger T_1F=diag [1,~e^{i2\pi/3},~e^{-i2\pi/3} ]$. The unitary matrix is not changed under such transformation. Therefore,  Eq.(6) can be rewritten in new bases $\textbf{V}^*_{i}=F^\dagger\textbf{V}_{i}$:
\begin{align}
\begin{split}
E^* \textbf{V}^*_{i}&=(\frac{3}{L_1}+\frac{1}{L_2}+\frac{6}{L_3}) \textbf{V}^*_{i}\\-
&\sum_{\langle j \rangle}  \frac{\rm I_{3\times 3}}{L_1}\textbf{V}^*_{j}-\sum_{\langle\langle k \rangle\rangle} \frac{T^*_1}{L_3} \textbf{V}^*_{k},
\end{split}
\end{align}
where $E^*=F^\dagger EF=\omega^2CT^*_0$. From above equation, $V_{i,1}^{*}$ is independent of $\omega$, which is reasonable to be discarded in the following . Thus, Eq.(8) can be divided into two independent equations,
\begin{align}
\begin{split}
3\omega^2C V^{*}_{i,2}&=(\frac{3}{L_1}+\frac{1}{L_2}+\frac{6}{L_3}) V^{*}_{i,2}- \\
&\sum_{\langle j \rangle} t V^{*}_{j,2}-\sum_{\langle\langle k \rangle\rangle}t_2e^{i2\pi/3} V^{*}_{k,2}.
\end{split}
\end{align}
\begin{align}
\begin{split}
3\omega^2C V^{*}_{i,3}&=(\frac{3}{L_1}+\frac{1}{L_2}+\frac{6}{L_3}) V^{*}_{i,3}- \\
&\sum_{\langle j \rangle} tV^{*}_{j,3}-\sum_{\langle\langle k \rangle\rangle}t_2e^{-i2\pi/3} V^{*}_{k,3}.
\end{split}
\end{align}
Eqs. (9) and (10) are nothing but the eigenequation for Haldane model.
If we regard $V^{*}_{i,2}$, $V^{*}_{i,3}$ as spin $\uparrow$ and $\downarrow$ wavefunction, the whole system resembles quantum spin Hall phase$\cite{Haldane,kanemele}$, where the corresponding spin component is decoupled Haldane model. Since two spin components are related by time reversal symmetry$\cite{LCrev1,u1h1,u1h2,LCkane}$, the topological properties of the system can be characterized by one of the spin  (i.e.  $V^{*}_{i,2}$ or $V^{*}_{i,3}$).

In the following, we only focus on the Haldane model with spin $S=\uparrow$ ($V^{*}_{i,2}$). From Eq. (9), the Haldane model has geometry phase $e^{\pm i2\pi/3}$. The energy $E_\uparrow=3\omega^2C$, nearest hopping $t=1/L_1$ and next nearest hopping $t_{2}=\frac{1}{L_3}$ can be determined by corresponding inductances and capacitances. More importantly, the on-site energy  $U_\uparrow=\frac{3}{L_1}+\frac{1}{L_2}+\frac{6}{L_3}$ has an independent parameter $L_2$. If one introduces the  randomization  to $L_2$, the on-site disorder can be achieved. This may pave a way on realization and detection  of exotic TAI in electric circuits. We note the Haldane model without $L_2$ and with other geometry phase was achieved in circuit experiment previously$\cite{u1h1}$. The $e^{ i2\pi/3}$ geometry phase breaks the particle-hole symmetry, which is benefitial to the topological Anderson transition.

\section{TAI in Haldane model}
In above section, we have demonstrated the construction of a special Haldane model in circuit. In this section, we study the general condition for realization of TAI in such Haldane model by considering the major characteristic of circuit (i.e. geometry phase $e^{ i2\pi/3}$ as well as positive inductance and capacitance value {\it etc}).

Eq.(9) is equivalent to tight-binding Hamiltonian$\cite{Haldane}$:
\begin{align}
\begin{split}
H&=-[\sum_{\langle i,j \rangle}t a^\dagger_ib_j+\sum_{\langle\langle i,k \rangle\rangle}t_{2}e^{i\frac{2\pi}{3}}(a^\dagger_ia_k+b^\dagger_ib_k)]\\
&
+\sum_i [(\Delta M_A+\epsilon_i )a^\dagger_ia_i+(\Delta M_B+\epsilon_i) b^\dagger_ib_i],
\end{split}
\end{align}
where $a_i, b_i$ are the annihilation operators of A and B sublattices. The first and second terms are nearest and next nearest hopping with  $t=1/L_1$ and $t_{2}=1/L_3$, respectively. For convenience, the hopping energy is fixed at $t=1$, $t_2=0.2t$ in the rest of this paper. The second line represents the on-site potential terms by $L_2$. We ignore $\frac{1}{L_1}$, $\frac{1}{L_3}$, because they only adjust the Fermi level and the topological nature is not affected. In order to obtain a disorder induced topological Anderson transition, the system should be a normal insulator in the clean limit. However, without stagger potential, the studied model is always topological insulator.  Fortunately, one can independently adjust $L_2$ for each site. Thus, not only a stagger potential for A/B sublattice, but also the Anderson disorder for each site can be introduced. Therefore, the on-site energy terms are separated into four sub-terms. $\Delta M_A$, $\Delta M_B$ with $\Delta =0.5$ indicates the on-site potential for A, B sublattice. $\epsilon_i$ is Anderson disorder$\cite{anderson}$ which is uniformly distributed in the range $[-\frac{W}{2},\frac{W}{2}]$ with the disorder strength $W$.
In real experiments, one can contact each sites with two grounding inductors in parallel to simulate the effect of $L_2$ for simplicity. That is to say, one inductor with value $\frac{1}{\Delta M_A} (\frac{1}{\Delta M_B})$ simulates the stagger potential, another inductor $L_W$ of random inductance simulates the disorder $\epsilon_i$.  The total on-site energy $\frac{1}{L_2} = \Delta M_{A/B} + \frac{1}{L_W}$. We note $L_W$ can capture
the major function of Anderson disorder, the general results in this section will not change even by considering their difference. We will discuss the similarity and the difference  in Sec.${\rm \uppercase\expandafter{\romannumeral4}}$.

The band structures for zigzag nano-ribbon with different $M_B$ are plotted in Fig.~\ref{band} [here, we call $3\omega^2C$ versus $k_x$ plots the band structure for simplicity]. When $M_B=M_A=t$, on-site potential of A and B sublattice are equal and the system is a topological insulator. As shown in Fig.~\ref{band}(a), the edge states appear in the bulk gap, while $\omega^2<0$ is not  considered for simplicity. In order to observe the topological Anderson transition, trivial states should be prepared firstly. Theoretically, it is impossible to realize negative on-site energy in circuits. That is to say, a real stagger potential $\Delta M_A=-\Delta M_B$ for A/B sublattice that was pointed out by Haldane $\cite{Haldane}$ can not be achieved in our system.  However, if $|\Delta M_A-\Delta M_B|$ is larger than the topological gap, a trivial state can also appear.  Therefore, a normal insulator is available even when $\Delta M_A>0$ and $\Delta M_B>0$.
A transition from topological insulator to normal insulator happens when $M_B$ is increased to $5t$, as plotted in Fig.~\ref{band} (b).

\begin{figure}
    \centering
	\includegraphics[width=0.5\textwidth]{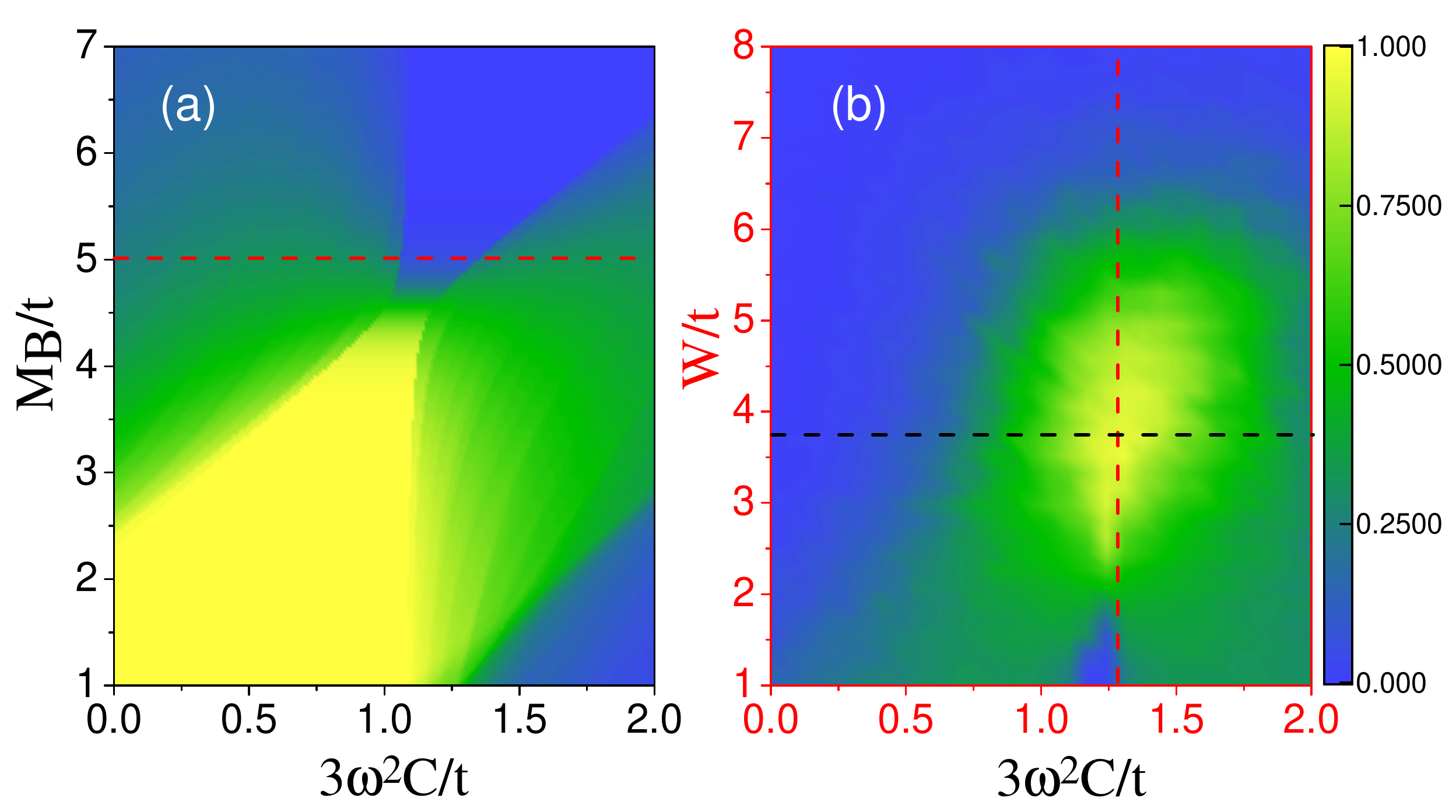}
	\caption{(Color online) (a) The Chern number versus  $M_B$ and  $3\omega^2C$. (b) The Chern number versus disorder strength $W$ and energy $3\omega^2C$ with $M_B=5t$. The size of the system is $N \times N=30 \times 30$. The other parameters are the same as those in Fig.~\ref{band}.}\label{chern}
\end{figure}

In order to make above topological transition more convincingly, the dependence of the Chern number on the energy $3\omega^2C$ for different $M_B$ is demonstrated in Fig.~\ref{chern}(a) with the help of non-commutative geometry method$\cite{song2,hug1,hug2,FDY1,FDY2,FDY3}$. For $M_B=t$, Chern number is quantized when $0<3\omega^2C<1.25t$, which is consistent with the spectrum shown in Fig.~\ref{band}(a). Once $3\omega^2C>1.25t$, bulk states become metallic and Chern number is no longer quantized. Moreover, the area of topological states is decreased with the increase of $M_B$ and topological phase totally disappears when $M_B\approx 4.5t$. In particular, the calculation reconfirms that $M_B=5t$ [marked by red dashed line in Fig.~\ref{chern}(a))] is a normal insulator. Such normal insulator is the starting point to pursue TAI in circuit.

Since non-commutative geometry method is based on real space wave function, we are able to obtain the evolution of Chern number of the Haldane model with the variation of disorder strength $W$. If Chern number becomes quantized under a fixed $W$ when $M_B=5t$, it enters into the TAI. As shown in Fig.~\ref{chern}(b), Chern number is almost quantized when $3t<W<5t$, $1.1t<3\omega^2C<1.5t$ for a $30 \times 30$ square sample. We also find the  quantized Chern number area is growing by increasing sample size $N$ due to the finite size effect$\cite{FinSize}$. Guiding by Fig.~\ref{chern}(b), we  propose the red dotted line case ($3\omega^2C\approx 1.3t$) as the best condition to realize TAI in circuits.

\begin{figure}
    \centering
	\includegraphics[width=0.45\textwidth]{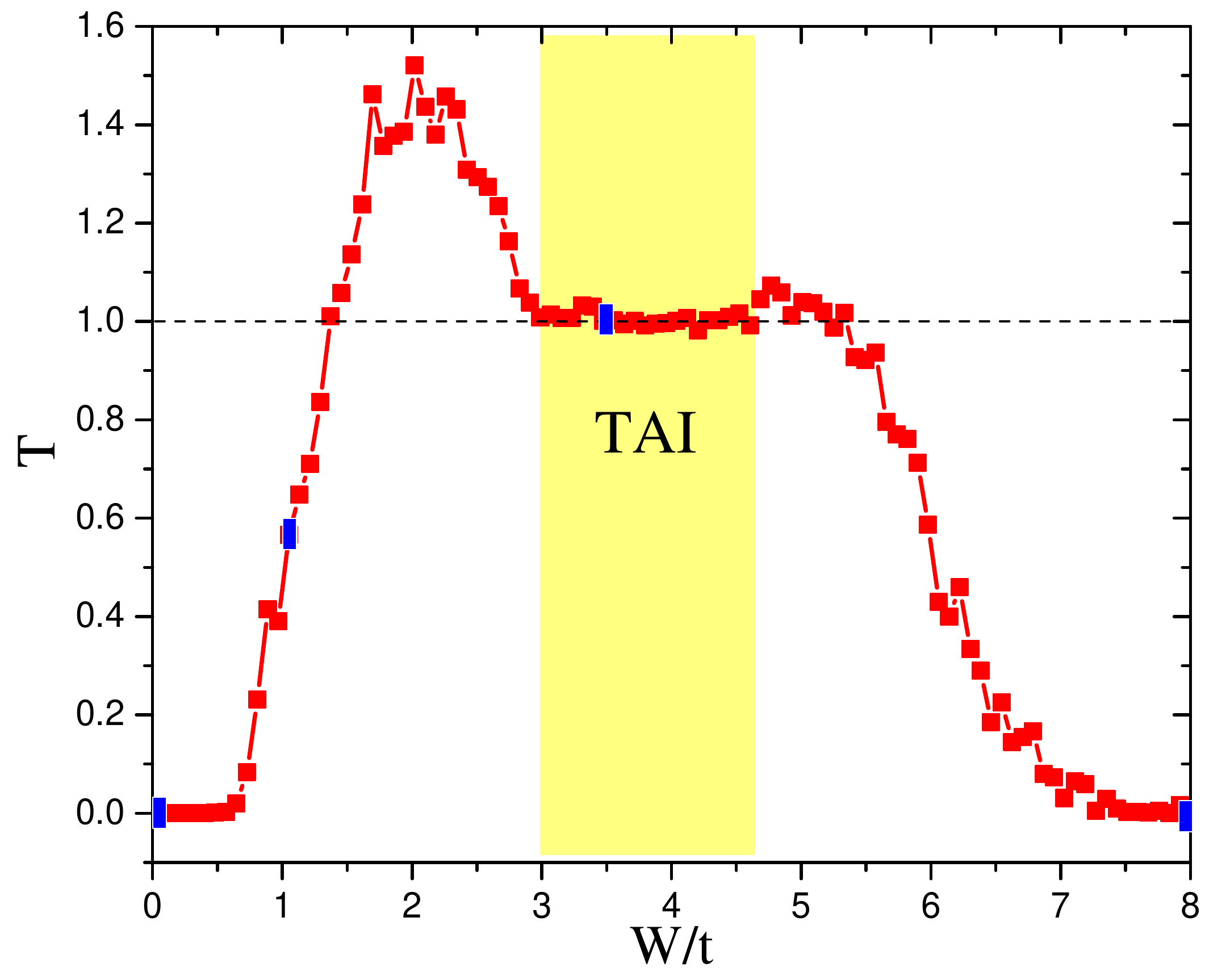}
	\caption{(Color online).  The transmission coefficient versus disorder strength $W$ with $M_B=5t,3\omega^2C=1.3t$ and $N=70$. White and yellow region indicate normal phase and TAI phase,respectively. The rectangles shows the disorder strength we used in Fig.~\ref{waveD}(a)-(d).  }\label{Trans}
\end{figure}

The transport simulation is also widely used to better characterize the TAI phase. It is known that the quantized transmission coefficient is equal to the number of edge states and the Chern number even for dirty samples.  The transmission coefficient versus disorder $W$ is obtained from nonequilibrium Green function method $\cite{datta,Hua1}$ and is plotted in Fig.~\ref{Trans}. For clean samples, the transmission coefficient is zero, which indicates its insulating properties. Then, the band structure is renormalized$\cite{Chen1,song1,lilun}$  with the increase of disorder strength $W$. If $W$ is large enough, bulk gap closes and the transmission coefficient is finite, but not quantized. However, for $3t<W<4.6t$, transmission coefficient becomes quantized, which clearly manifests the disorder induced topological Anderson transition (filled with yellow in Fig.~\ref{Trans}). If one continues to increase $W$, all states are localized with decreasing of transmission coefficient finally. These results are well consistent with Fig.~\ref{chern}(b).


\section{Comprehensive understanding of disorder in circuit}
In section III, based on Anderson disorder simulation in Haldane model, we verify that TAI phase can in principle exist in electric circuit. However, the disorder (randomness of inductor) in circuit may posses its own characteristic, making its consequence more or less different from Anderson disorder. In this section, we present a comprehensive study.

Anderson disorder is uniformly distributed in the range $[-\frac{W}{2},\frac{W}{2}]$ with disorder strength $W$. Because
negative on-site potential is not accessible in circuits, we will design some alternative scheme to achieve the function of Anderson disorder in such system. We also compare the scheme with standard Anderson disorder. Suppose node $\alpha$ is grounded by a inductor with value $L_W$, on-site potential $U_{\alpha \uparrow}$ will be added by $1/L_W$. If different nodes are grounded by the inductor with different inductance $L_W$, disorder can be realized. The first scheme is to choose inductor with value between $1/W$ and $\infty$ to let the on-site potential in the range $[0,W]$. The case discussed above is equivalent to that A/B sublattice potential is changed to $ \Delta M_{A/B} + W/2$ and disorder is still in the range  $[-\frac{W}{2},\frac{W}{2}]$. Since initial topological index of the system is still determined by stagger potential $|\Delta M_A-\Delta M_B|$, this scheme only shifts Fermi energy but not changes the topological Anderson transition.  The scheme can simulate Anderson disorder well. However, the value of inductors should be carefully picked to achieve a uniform distribution of on-site potential and inductors with huge value $L_W  \approx \frac{1}{0} = \infty$ will be needed. Although this scheme is feasible in principle, it is hard for experimental implementation.

\begin{figure}
    \centering
	\includegraphics[width=0.45\textwidth]{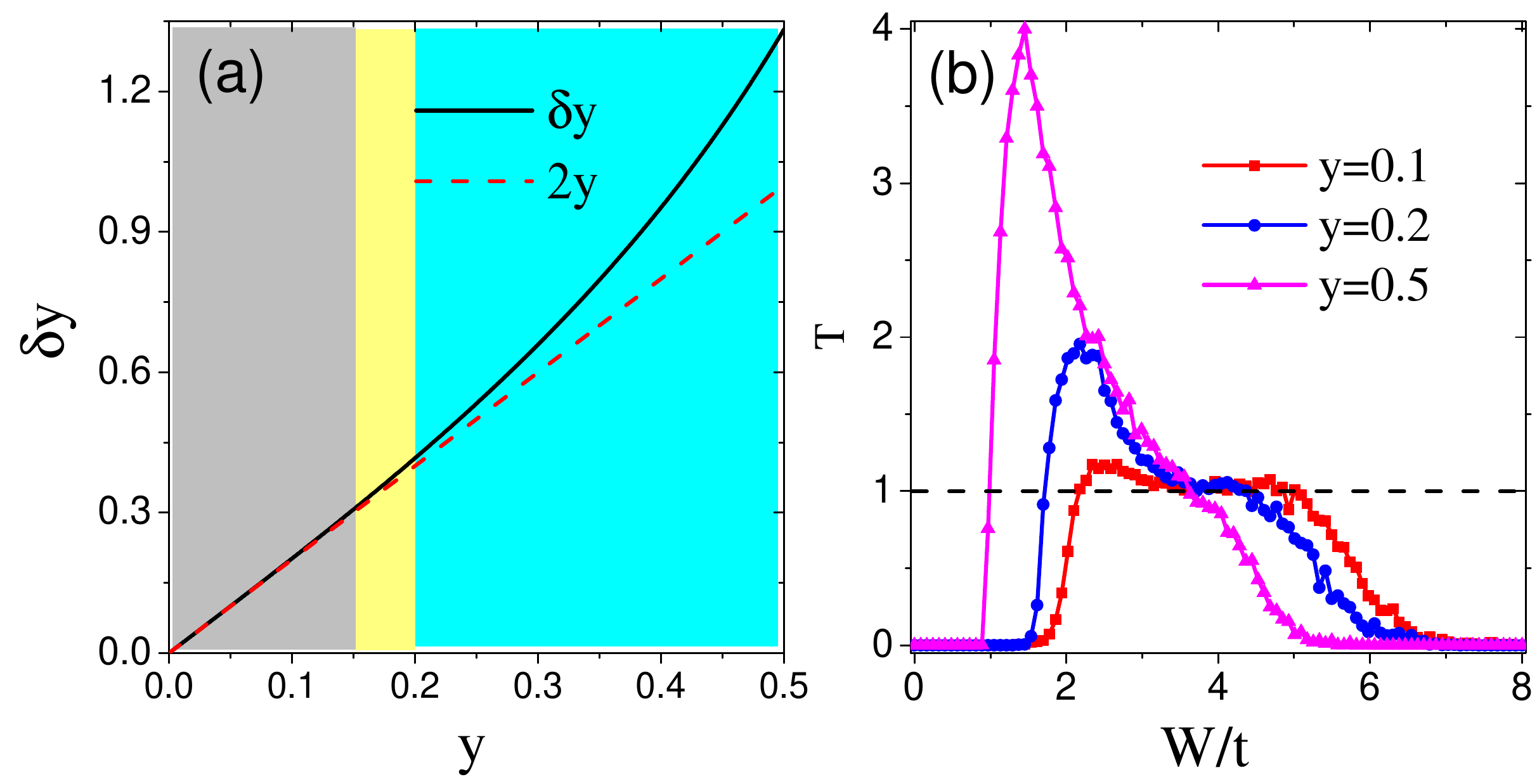}
	\caption{(Color online). (a) The difference between the maximum and minimum value disorders vs error ratio $y$. The black line and red line indicate $\delta y$ (inductance error) and $2y$ (Anderson disorder), respectively. (b) The transmission coefficient versus $W$ for different error ratio. The other parameters are taken the same as those of Fig.~\ref{Trans}. }\label{prop}
\end{figure}

Another scheme is to make use of the error ratio $y$ of inductors with inductance $L_W$. We hope randomness of inductor could induce a disorder that like Anderson type disorder. Let the accurate inductance $L_W=X$. Due to the fluctuation, the realistic inductance is in between $X(1\pm y)$$\cite{LCHO1}$. Thus, the difference between the maximum and minimum on-site potential induced by  error ratio $y$ is $\delta y=\frac{1}{X}(\frac{1}{1-y}-\frac{1}{1+y})$. The Fermi energy of $3\omega^2C$ is shifted up to $\frac{1}{X}(\frac{1}{1-y}+\frac{1}{1+y})/2$, which has no influence on topological properties.
To compare random inductance induced disorder with Anderson disorder, we plot the evolution of $\delta y$ with the increase of $y$ in Fig.~\ref{prop}(a). The red dashed line represents the standard Anderson disorder with uniform distribution $[-y,y]$ and the difference between the maximum and minimum value is $2y$. The black line represents the random inductance induced disorder with $X=1$. Because the black line and the red dashed line are completely coincident when $y<0.15$, the random inductance induced disorder can simulate Anderson disorder quite well. Random inductance induced disorder deviates from Anderson disorder when $0.15<y<0.2$. When $y>0.2$, $\delta y$ and $2y$ show sharp difference behavior . Therefore, random inductance induced disorder and Anderson disorder have totally  different distributions.

In order to find the differences among these three inductance error ratio regions marked in Fig.~\ref{prop}(a), we also study the influence of $y$ with the same disorder strength $W$. Firstly, we generate an inductance $X$ with its error unit distribution in the range $E_y \in [-y,y]$. The on site potential $E_U$ cause by $E_y$ is
\begin{equation}
E_U=\frac{1}{X}\frac{1}{1+E_y}-\frac{1}{2X}(\frac{1}{1+y}+\frac{1}{1-y}).
\end{equation}
\begin{figure}
    \centering
	\includegraphics[width=0.45\textwidth]{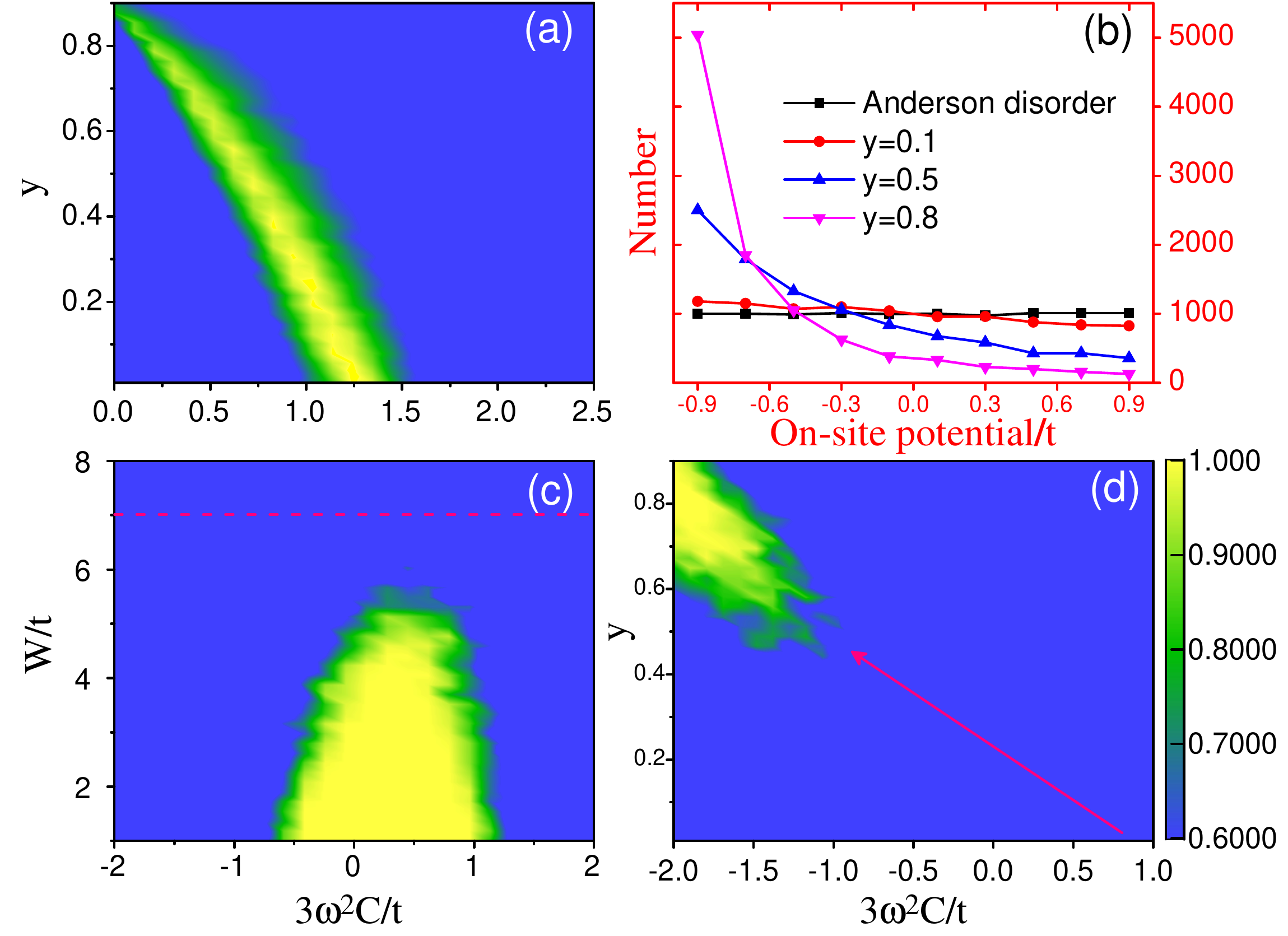}
	\caption{(Color online).  (a).The Chern number versus error ratio $y$ and $3\omega^2C$. Disorder strength is $W=3.5t$ and other parameters are the same as those of Fig.~\ref{chern}(b). (b) demonstrates the distribution of Anderson disorder and random inductance induced disorder $E_U$ for different $y$ with $W=2t$. (c).The evolution of Chern number with the increase of Anderson disorder strength $W$ for $M_B$=t. Other parameters are the same with Fig.~\ref{chern}(a). The red dash line marks the position of $W=7t$. (d) The evolution of Chern number with the increase of $y$ for $E_U$ and disorder strength is fixed at $W=7t$[ marked in Fig.~\ref{spe}(c)]. Other parameters are the same with (c).}\label{spe}
\end{figure}
Here, $\frac{1}{2X}(\frac{1}{1+y}+\frac{1}{1-y})$ is the
shift of Fermi energy, which should be substracted. From above paragraph, the on-site potential difference $\delta y$ defines the disorder strength  as $W=\frac{1}{X}(\frac{1}{1-y}-\frac{1}{1+y})$.
Then, by substituting $X$ with $W$ in Eq. (12), one obtain the on-site potential induced by $L_W$
\begin{equation}
E_{U}=\frac{W(1-y^2)}{2y}(\frac{1}{1+E_y}-\frac{1}{1-y^2}).
\end{equation}
From Eq. (13), we not only obtain the on-site potential caused by inductance error in circuit Haldane model, but also can compare the effect of different $y$ with the same  $W$.

The evolution of transmission coefficient for different value of $y$ ($y=0.1$, $0.2$, $0.5$) is plotted in Fig.~\ref{prop}(b). For $y=0.1$, the inaccurate inductance induced disorder $E_U$ can simulate Anderson disorder well. The transmission coefficient is quantized when $3t<W<4.8t$, which is almost the same as the results in Fig.~\ref{Trans}. Such result once again verifies that TAI phase can be obtained in the frame of circuit system.  For $y=0.2$, $E_U$ deviates from Anderson type disorder. Although the transmission coefficient  $T=1$ plateau still exists, the plateau width shrinks greatly. However, $E_U$ is totally different from Anderson disorder for $y=0.5$ [see Fig.~\ref{prop}(a)]. The $T=1$ plateau disappears, which means the TAI is not available.

Now, we fix disorder strength $W=3.5t$ to study how error ratio $y$ affects topological Anderson phase [see Fig.~\ref{spe}(a)].  The topological region is gradually deflected toward the low energy direction with the increase of $y$.
Remarkably, we also find, by increasing $y$, the energy window for TAI is gradually decreased. And it disappears when $y$ approaches $0.8$.
In order to understand above phenomenon, we calculate the distribution of $E_U$, as shown in Fig.~\ref{spe}(b). For the fixed disorder strength $W=2t$, we randomly generate 10000 numbers of $E_U$ between $-t$ and $t$ by Eq.(13)  and divide them into $10$ equal intervals [i.e. $E_U \in [-t,-0.8t]$]. Then the frequency that appears in each interval range is counted, as shown in Fig.~\ref{spe}(b).
The statistics of Anderson disorder, which is uniformly distributed,  is also plotted for comparison. For $y=0.1$, random inductance induced disorder almost agrees with Anderson disorder, which is consistent with the previous analysis. With the increase of $y$, the curves gradually deviates from Anderson disorder. Interestingly, the counting number is greatly increased for small $W$. Since $E_U$ is not symmetric about the origin, the topological region deflects toward the low energy direction for TAI [see Fig.~\ref{spe}(a)]. Moreover, as shown in Fig.~\ref{spe}(b), the on-site potential is mainly concentrated in a small range for $y=0.8$ [e.g. $E_U \in [-t,-0.8t]$]. In this case, $E_U$ plays a role as weak disorder rather than strong disorder. It can greatly shift the Fermi energy but has little effect on the renormalization of topological mass from positive to negative $\cite{lilun}$.
Consequently, the sample tends to become clean, and thus topological Anderson phase is gone.

To confirm such speculation, we also study the influence of $y$ on the transition from Chern insulator to Anderson insulator  caused by strong disorder. In Fig.~\ref{spe}(c), the evolution of Chern number under Anderson disorder is plotted. We set $M_B=t$ and the clean system belongs to topological insulator. For $W \approx 7t$ [see red line in Fig.~\ref{spe}(c)], the Anderson disorder drives the system into the Anderson insulator phase and a zero Chern number is obtained. Then, we fix disorder strength $W \approx 7t$ but replace the Anderson disorder with the random inductance induced disorder $E_U$. Fig.~\ref{spe}(d) plots evolution of Chern number under different $y$. Topological states reappear with large $y$. This shows strong evidence that, even for fixed disorder strength $W$,  the disorder effect of  $E_U$ becomes weaker and weaker by increasing $y$. Finally, the sample is close to the clean sample in large $y$ limit.

Conclusively, the random inductance induced disorder in circuit plays the same role as Anderson disorder when error ratio $y$ is small. However, the random inductance induced disorder and Anderson disorder show different behaviors when $y$ is larger. Ultimately,  the disordered circuit with large enough $y$ behaves like a clean system, where the topological Anderson transition cannot be achieved.

\section{Experimental detection of TAI in circuits}
Thus far, based on Haldane model in electric circuits, we have confirmed the existence of TAI. The well developed electrical techniques not only make the construct of TAI in circuit feasible, but also provide a simple way to detect such phase. In the following, we demonstrate two methods to detect TAI phase on account of circuit characteristics.

The first method is based on Green's function. Different from condensed matter system, Green's function is the impedance$\cite{LCHO1}$ in circuits, which can be directly measured. From Kirchhoff's current law, we obtain
\begin{equation}
\begin{array}{l}
\textbf{I}^*=i\omega^{-1}(3\omega^2C-H)\textbf{V}^*\\
G\textbf{I}^*=i\omega^{-1}\textbf{V}^*,\\
\end{array}
\end{equation}
 with $G=\frac{1}{3\omega^2C-H}$ and $H$ is the Hamiltonian. If there is an input current $I_0$ at node $n$, the voltage of node $m$ satisfies $V_m=-i\omega G_{mn}I_0$ and Green's function $G_{mn}$ can be obtained$\cite{LCgreen}$. Therefore, for a square sample with $N^2$ sites, the entire Green's function $G$ can be detected through $N^2$ times input of node current. Then, the Hamiltonian is $H=3\omega^2C-G^{-1}$ and all the information of dirty sample can be achieved. Interestingly, the detection of transmission coefficient is much easier in experiment. According to previous theory$\cite{datta,PNjunction}$, only the Green's function $g_c$ between the $1st$  and the $N_{th}$ principle line
\begin{equation}
g_c=\left(
\begin{array}{cc}
g_{11} & g_{1N}\\
g_{N1} & g_{NN}\\
\end{array}
\right),
\end{equation}
is need. Here, the left/right semi-infinite lead [source and drain] information is not measured. However, since there is no disorder, the retarded self-energy $\Sigma^r_{L/R}$ of these two leads can be numerically$\cite{zineng1,zineng2}$ obtained. Then, one gets retarded Green's function of sample according to Dyson's equation$\cite{Dyson}$ $g_c^r=(g_c^{-1}-\Sigma^r_L-\Sigma_R^r)^{-1}$. Finally, with the help of non-equilibrium Green's function method$\cite{datta}$, the transmission coefficient is expressed as $T=Tr(\Gamma_Lg^r_c\Gamma_Rg^a_c)$ with $\Gamma_{L/R}=i(\Sigma^r_{L/R}-\Sigma^a_{L/R})$. As shown in Fig.~\ref{Trans}, transmission coefficient should be quantized ($T=1$) once there is a topological Anderson transition.
Nevertheless, Fig.~\ref{Trans} is difficult to be obtained experimentally since it is hard to frequently change disorder strength $W$ in circuit. Fortunately, the feature of TAI can be confirmed from  transmission coefficient $T$ with  twice measurements. Because, one can easily measure
the evolution of $T$ for different energy $3\omega^2C$ in a clean sample ($W=0$) and the transmission coefficient for a sample with fixed disorder strength $W$. $T$ will be quantized to one (zero) in a scope of $3\omega^2C$ for TAI (normal insulator), as shown in black line in Fig.~\ref{chern}(b) [red line in Fig.~\ref{chern}(a)].

We emphasize in Eq. (14), the effective current $\textbf{I}^*$ is not a direct input quantity.  In our  circuit model, the real input current $\textbf{I}_{i}$ for site $i$ should satisfy $\textbf{I}^*_{i}=F^\dagger\textbf{I}_{i}$. If $\textbf{I}_{i}=(I_1,I_2,I_3)$ meets the following relations
\begin{equation}
\begin{array}{l}
I_1+I_2+I_3=0\\
I_1+e^{-i2\pi/3}I_2+e^{-i4\pi/3}I_3=\sqrt{3}I_0\\
I_1+e^{-i4\pi/3}I_2+e^{-i8\pi/3}I_3=0,\\
\end{array}
\end{equation}
one obtain $\textbf{I}^*_{i}=(0,I_0,0)$. This means only  spin up modes of Haldane model (Eq.(9)) are excited. And from Eq.(14), one can obtain
the Green's function and Hamiltonian for spin up Haldane model.
\begin{figure}
    \centering
	\includegraphics[width=0.45\textwidth]{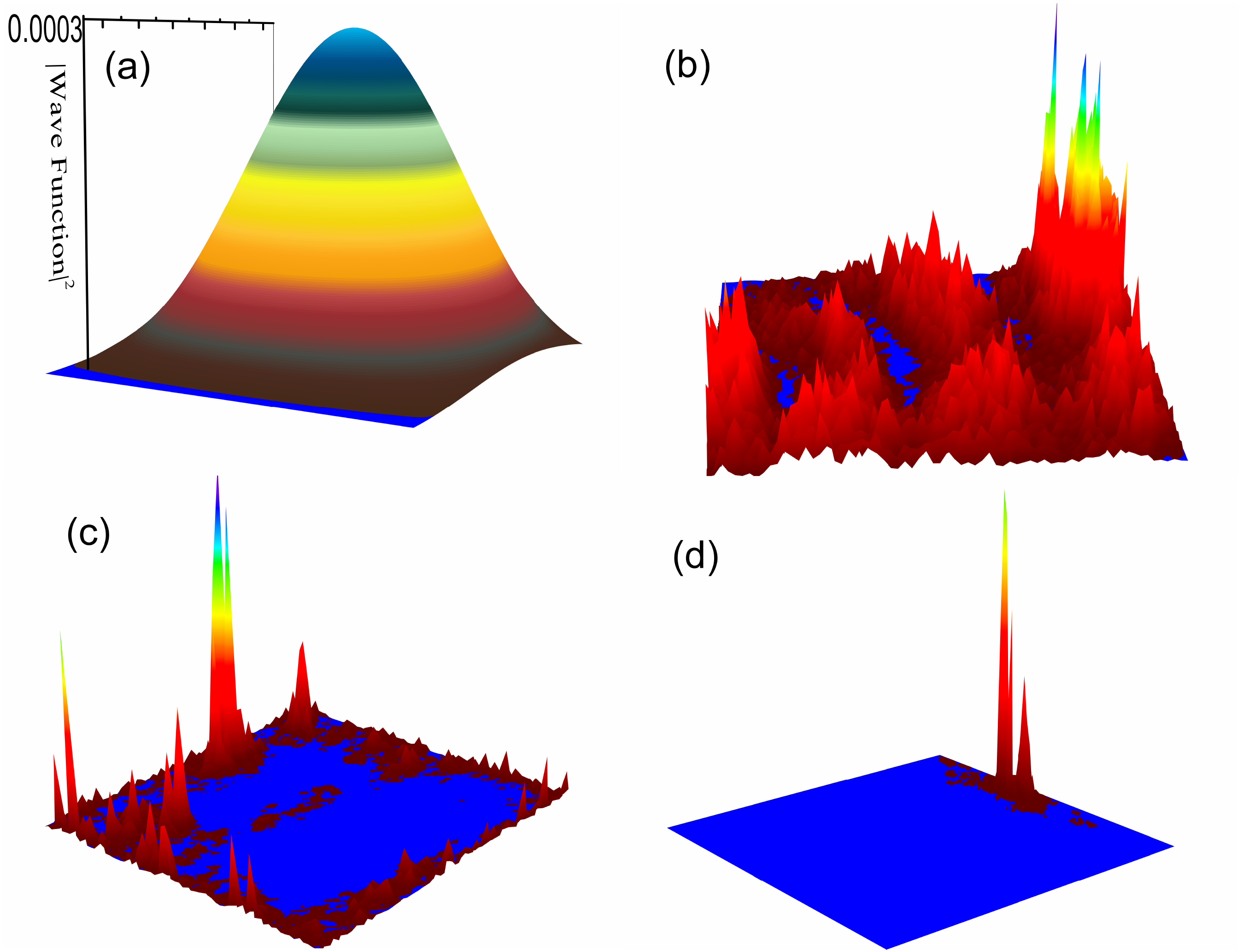}
	\caption{(Color online).  The distribution of wave function $|V|^2$ for different disorder strength $W$. (a) $W=0$, (b) $W=1.5t$, (c) $W=3.5t$ and (d) $W=8t$. The value of $W$ which used for this plot are marked in Fig.~\ref{Trans} by rectangles. The other parameters are the same as those of Fig.~\ref{Trans}.}\label{waveD}
\end{figure}
\begin{figure}
    \centering
	\includegraphics[width=0.45\textwidth]{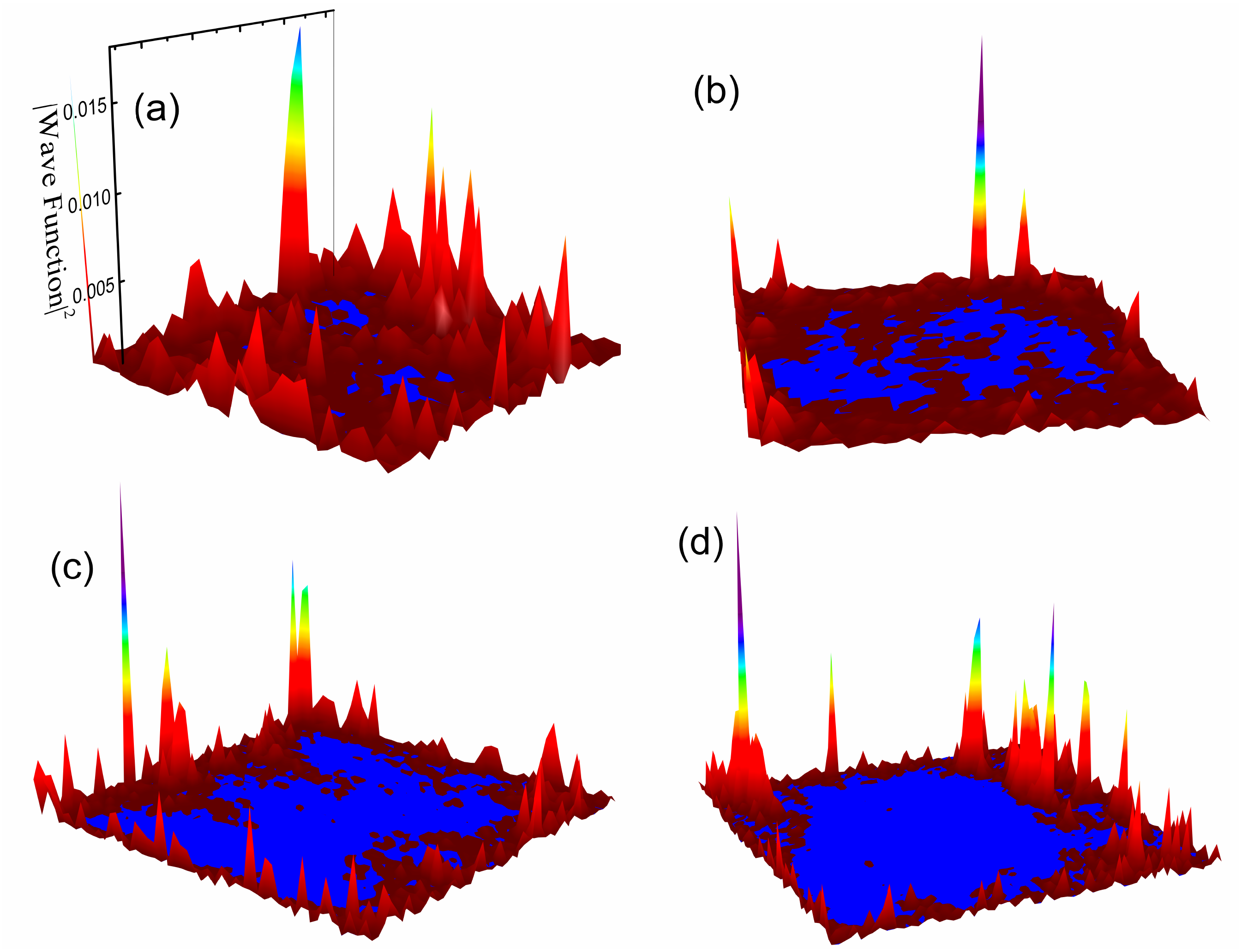}
	\caption{(Color online).  The wave function distribution $|V|^2$ for square samples $N\times N$ with different width $N$. (a) $N=30$, (b) $N=40$, (c) $N=50$ and (d) $N=60$. The other parameters are the same as those of Fig.~\ref{waveD}(c). }\label{waveN}
\end{figure}

Another feasible method is to measure the spatial distribution of wave function. As stated in Sec. II, voltages in circuits correspond to wave functions in condensed matter physics. Experimentally, voltage is much easier to be measured by voltmeter and it has been widely used in topological state  studies of circuit  $\cite{LCrev2,u1h1}$. Therefore, if an inductor is excited, the propagation of the excitation makes wave function measurable$\cite{LCedge}$. Thus, the detection of edge states  under different disorder can be used to  verify the existence of TAI.
The evolution of wave function under disorder strength $W=0$, $1.5t$, $3.5t$, $8t$ (marked in Fig.~\ref{Trans} by rectangles)
is plotted in Fig.~\ref{waveD}.  Since it is a normal insulator for clean sample, the wave function is mainly located on the center of the sample [see Fig.~\ref{waveD} (a)]. With increasing $W$, the eigen-state spreads to the whole sample and belongs to metallic state for $W=1.5t$. When $W=3.5t$, the wave function distributes mainly on the edge, which indicates the topological nature of TAI. In the end, all states are totally localized for $W=8t$ and the transmission coefficient is zero because of Anderson localization. In order to make the signal of TAI more clear,  the examination of the wave function on the sample size  is also suggested.  Fig.~\ref{waveN} shows distribution of wave function under different size $N$ at $W=3.5t$. When $N$ is too small, e.g. $N=30$, the edge states belong to opposite edges will couple  with each other due to finite size effect, making the TAI undetectable. However, for large $N$, the distribution of edge state becomes obvious and insensitive to $N$. So, it is very easy to be detected when $N\ge 50$. The  size  simulation not only provides a way to find the appropriate experimental condition for TAI, but also reveals the topological nature of TAI.

Specially, different from cold atom systems and  photonic crystal systems, where either edge states or quantizied transmission coefficient is difficult to be observed, one can detect these two quantities simultaneously in circuit.

\section{Discussion and Conclusion}
Finally, we propose the appropriate parameters and system sizes for realizing TAI in detail. Take $t$ as the unit. The nearest and next nearest hopping can be realized with inductance set as $L_1=\frac{1}{t}$ and $L_3=\frac{5}{t}$. Furthermore, the value of accurate grounding inductors, which stimulate the stagger potential for A/B sublattice, are $\frac{1}{\Delta M_A}=\frac{2}{t}$ and $\frac{1}{\Delta M_B}=\frac{0.4}{t}$, respectively. In addition, Anderson disorder with strength $W\approx 3.5t$ can be achieved  by inductance $L_W\approx \frac{0.0577}{t}$ with error ratio $y=0.1$ and  the Fermi energy  shift is approximately $ 17.5t$. Therefore, the best Fermi energy by adding omit shift $\frac{3}{L_1} +\frac{6}{L_3} + 17.5t $ is $3\omega^2C\approx 23 t$, with $\omega$ denotes input frequency and $C$ denotes capacitance.
Experimentally, if we set $L_1=1~{\rm mH}$, $L_3=5~{\rm mH}$ and $C=10~{\rm \mu F}$, the other parameters are  as follows. The accurate grounding inductors for A/B sublattice are $2~{\rm mH}$ and $0.4~{\rm mH}$, respectively. The disordered grounding inductors have the value ${L_W \approx 57~{\rm \mu H}}$ and input frequency for current or voltage is $\omega\approx 27.7~{\rm MHz}$ with system size $N \geq 50$.

In summary, we present a scheme for implementing TAI in electric circuits by constructing disordered Haldane model. With the help of Chern number calculation, we demonstrate that TAI can be realized through random inductance induced disorder $E_U$. However, the topological Anderson transition in circuit poses its own properites. For small inductance error ratio $y$, $E_U$ and Anderson disorder have similar distribution, the topological Anderson phase can be obtained. For larger $y$, the two kinds of disorder show a sharp contrast in distribution, the topological Anderson phase will be not available. Finally, based on the measurement of wave function and transmission coefficient in circuit, two experiments are suggested to detect TAI. Specially, due to the nontrivial property, TAI can be  fabricated and be measured even with only one dirty circuit sample.

\section{ACKNOWLEDGMENTS}  We are grateful to R. Yu, C. Z. Chen and Q. J. Wang for helpful discussion. This work was supported by NSFC under Grant Nos. 11534001, 11822407, 11874139, NSF of Jiangsu Province under Grants Nos. BK2016007 and a Project Funded by the Priority Academic Program Development of Jiangsu Higher Education Institutions (PAPD).

\end{document}